\documentclass[12pt]{article}

\usepackage{amsmath,amsfonts,amssymb}
\usepackage{graphicx}
\usepackage{psfrag}
\usepackage{enumerate}
\usepackage{mathrsfs}

\makeatletter\renewcommand\section{\@startsection {section}{1}{\z@}%
                                   {-3.5ex \@plus -1ex \@minus -.2ex}%nn
                                   {2.3ex \@plus.2ex}%
                                   {\normalfont\large\bfseries}}
\renewcommand\subsection{\@startsection{subsection}{2}{\z@}%
                                     {-3.25ex\@plus -1ex \@minus -.2ex}%
                                     {1.5ex \@plus .2ex}%
                                     {\normalfont\bfseries}}

\newcommand{\be}{\begin{equation}}
\newcommand{\ee}{\end{equation}}

\newcommand{\eeq}{\end{eqnarray}}

\def\[{\left [}
\def\]{\right ]}
\def\({\left (}
\def\){\right )}

\def\r2{\sqrt{2}}

\def\Tr{{\rm Tr}}

\newcommand{\bbibitem}[1]{\bibitem{#1}\marginpar{#1}}

\def\Label#1{\label{#1}%
  \smash{\hbox to0pt{\raise1ex\hbox{\tiny[#1]}\hss}}}
\def\noLabels{\let\Label=\label}
\def\nobbibitem{\let\bbibitem=\bibitem}

\begin{document}
\noLabels % uncomment for final production
\nobbibitem % uncomment for final production

%\begin{titlepage}

%\begin{flushright}%\vspace{-2cm}
%{\small
%UPR-1154-T  \\ %\vspace{-0.35cm}
%LBNL-60486 \\
%hep-th/0606118}%\\
%\end{flushright}
%\vspace{12 mm}

%\vfil\
%vfil

\begin{center}

%\begin{flushright} \vspace{-3cm}
%{\small UPR-1222-T}  \\
%\end{flushright}
%\vspace{2cm}

%{\Large \bf Entropy of spherical Rindler space}
{\Large \bf  The entropy of a hole in spacetime}

\vspace{7mm}

Vijay Balasubramanian$^{a,b}$, Borun D. Chowdhury, 
Bart{\l}omiej Czech, Jan de Boer$^{c,}$\footnote{\tt email: vijay@physics.upenn.edu, czech@stanford.edu, bdchowdh@asu.edu, J.deBoer@uva.nl}
%\\

\vspace{5mm}

%\bigskip\centerline{$^a$\it Department of Physics and
%Astronomy}
\bigskip\centerline{$^a$\it David Rittenhouse Laboratories, University of Pennsylvania}
\smallskip\centerline{\it 209 S 33$^{\rm rd}$ Street, Philadelphia, PA 19104, USA}
\bigskip\medskip
\bigskip\centerline{$^b$\it Laboratoire de Physics Th\'{e}orique, \'{E}cole Normale Sup\'{e}rieure}
\smallskip\centerline{\it 24 rue Lhomond, 75005 Paris, France}
\bigskip\medskip
\bigskip\centerline{$^c$\it Institute for Theoretical Physics, University of Amsterdam}
\smallskip\centerline{\it Science Park 904, Postbus 94485, 1090 GL Amsterdam, The Netherlands}
\bigskip\medskip
%\vfil

\end{center}
\setcounter{footnote}{0}
\begin{abstract}
\noindent
We compute the gravitational entropy of  ``spherical Rindler space'', a time-dependent, spherically symmetric generalization of ordinary Rindler space, defined with reference to a family of observers traveling along non-parallel, accelerated trajectories.  All these observers are causally disconnected from a spherical region $H$  (a ``hole'') located at the origin of Minkowski space. The entropy evaluates to $S = \mathcal{A}/4G$, where $\mathcal{A}$ is the area of the spherical acceleration horizon, which coincides with the 
boundary of $H$. We propose that $S$ is the entropy of entanglement between quantum gravitational degrees of freedom supporting the interior and the exterior of the sphere $H$. 
\end{abstract}
\newpage
%\tableofcontents
%\section{Introduction}
%\label{intro}

\section{Introduction}

\begin{figure}
\centering
\includegraphics[keepaspectratio,width=0.7\linewidth]{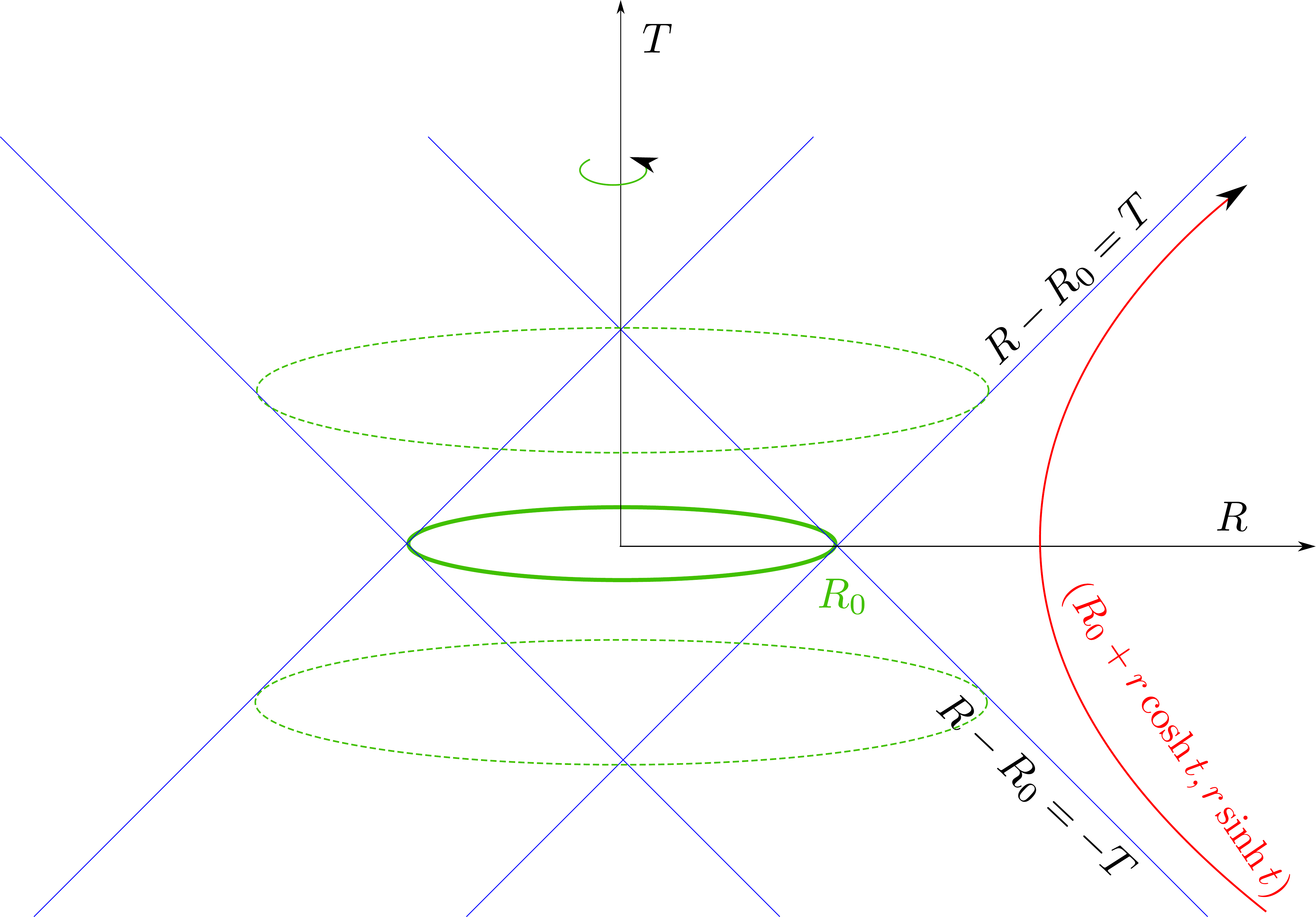}
\caption{Spherical Rindler Space
}    
\label{fig:sph}
\end{figure}

Ordinary Rindler space consists of the spacetime points from which signals can be exchanged with a uniformly accelerating observer in Minkowski space.   Here we study spherical Rindler space, which consists of spacetime points that can exchange signals with at least one out of a family of radially accelerating observers. All these observers are causally disconnected from a spherical region $H$ (a ``hole'') of radius $R_0$ located at the origin of Minkowski space (Fig.~\ref{fig:sph}).  The boundary of $H$ is a horizon.     The thermodynamics of this horizon is subtle, as spherical Rindler space is time-dependent for global reasons --  the observers who define it accelerate in different directions.   Thus, to define and compute the entropy of  this spacetime we develop a novel approach, which should extend  to a wider class of time-dependent universes.

Spherical Rindler space is relevant to understanding how spacetime arises from microscopic degrees of freedom.   To see this, recall first that Ryu and Takayanagi have proposed a relation between areas of extremal surfaces in asymptotically anti-de Sitter spaces and quantum entanglement in a dual field theory \cite{rt} (the time-dependent generalization is in \cite{hrt}).   Extending this idea, Van Raamsdonk has suggested that connectedness in spacetime arises from entanglement of the underlying quantum gravitational degrees of freedom \cite{mav1, mav2}. In some situations, reducing the entanglement between two gravitating systems can be interpreted as dissecting an otherwise connected spacetime into two disjoint components \cite{rqg}, each of which ends on a singularity resembling a black hole firewall \cite{pre-amps, firewalls}.  Is it possible to similarly dissect the interior and exterior of a spherical ball in spacetime? In the Discussion below we put forward a simple argument explaining why the computations presented in this paper measure the entanglement entropy between the quantum gravity systems interior and exterior to a spherical ball in flat space.  If we lift our computation to anti-de Sitter space by introducing a tiny negative cosmological constant, this entropy becomes holographically related to the entanglement between  ultraviolet and  infrared sectors of a dual field theory \cite{uvir}. We expect this type of entanglement across scales to be a necessary condition for a quantum system to have a gravitational dual.

When we were preparing this manuscript, a related paper \cite{maldacena} appeared.  We comment on its relation with our work in the Discussion.

\section{Spherical Rindler space}

Consider $(d+1)$-dimensional Minkowski space:
\begin{equation}
ds^2 = -dT^2 + dX^2 + d\vec{Y}_{d-1}^2
\end{equation}
An accelerated observer can exchange signals with a subregion of the spacetime called Rindler space. A set of coordinates covering Rindler space is given by:
\begin{equation}
x = \sqrt{(X - X_0)^2 - T^2} \qquad {\rm and} \qquad t = \tanh^{-1}\frac{T}{X-X_0}
\end{equation}
With these definitions, the metric takes the form
\begin{equation}
ds^2 = -x^2 dt^2 + dx^2 + d\vec{Y}_{d-1}^2\,,  \label{planar}
\end{equation}
with $x=0$ marking the Rindler horizon.

We are interested in a generalization of Rindler space, which is appropriate for a family of observers accelerating away from a common center, who are causally disconnected from a spherical region of radius $R_0$. Starting again with Minkowski space, now in spherical coordinates, 
\begin{equation}
ds^2 = -dT^2 + dR^2 + R^2 d\Omega_{d-1}^2\,, \label{radial}
\end{equation}
define spherical Rindler coordinates:
\begin{equation}
r = \sqrt{(R-R_0)^2 - T^2} \qquad {\rm and} \qquad t = \tanh^{-1}\frac{T}{R-R_0} 
\label{sphrindler}
\end{equation}
These coordinates cover the region of Minkowski space from which signals can be exchanged with at least one observer out of a family of observers accelerating in the radial direction. This region ends on a horizon, because none of our accelerated observers can see the inside of a sphere of radius $R_0$ at the center of Minkowski space. The horizon is by construction spherically symmetric and its size is also given by $R_0$. The metric takes the form:
\begin{equation}
ds^2 = -r^2 dt^2 + dr^2 + (R_0 + r \cosh t)^2 d\Omega_{d-1}^2 \label{sph}
\end{equation}
We shall refer to this geometry as spherical Rindler space.   Its Euclidean continuation is:
\begin{equation}
ds_E^2 = r^2 d\tau^2 + dr^2 + (R_0 + r \cos \tau)^2 d\Omega_{d-1}^2  \label{Esph}
\end{equation}
Regularity of this metric at $r=0$ and single-valuedness over the sphere require that this metric be periodic in imaginary time $\tau \sim \tau + 2\pi$.  Thus, the temporal circle pinches off smoothly at $r=0$, just as it does for Euclidean  planar Rindler and black hole spaces.

It is well known that planar Rindler space (\ref{planar}) has a gravitational entropy given by the area of the horizon divided by $4G$ \cite{laflamme}. This entropy,  computed from the Einstein-Hilbert action, is purely gravitational. Our goal is to compute the gravitational entropy of spherical Rindler space (\ref{sph}). A novel feature here is the time-dependence of the metric.    This means that one must be careful when trying to apply formalisms, which were successful in the conventional, static cases of planar Rindler and black hole spacetimes.

\section{Standard techniques} 
\label{stand}

We will recall several standard methods of computing the entropy of static gravitating spacetimes (see 
  \cite{bekenstein, bch, gibbonshawking, centennial, laflamme, kallosh, susskind, iyerwald, iyerwald2, Parattu:2013gwa} and the review \cite{simonreview}).   The general idea is to interpret the Euclidean path integral of gravity with fixed charges and mass as a partition function, so that the action evaluated on saddlepoints gives a semiclassical approximation to the free energy.    In these saddlepoints, time is compact and interpreted as a thermal circle, which typically closes off at a point  corresponding to the location of the Lorentzian horizon, leading to a ``cigar'' geometry.     Typically, we have a $U(1)$ invariance around the Eulidean time, indicating a system in thermal equilibrium.     Regularity of the  saddlepoint imposes a specific periodicity on Euclidean time, thus relating the global charges and temperature (e.g., $\beta = 8\pi M$ when $M$ is the mass of a 4d Schwarzschild black hole).   If we consider solutions with a different periodicity on the thermal circle (i.e., treating mass and temperature as independent variables for a black hole,) there will be a conical defect at the tip of the ``cigar'' where the thermal circle closes off.    There are now several ways to compute the entropy in terms of these geometries.

\paragraph{(a) Direct computation of the entropy: }  For a system in equilibrium with energy $M$, free energy $F$, and inverse temperature $\beta$,   there is a thermodynamic relation for the entropy:
\begin{equation}
S = \beta M - \beta F
\label{Seql}
\end{equation}
To use this equation in gravity, we interpret the Euclidean gravitational action  as computing the free energy ($I_{\rm gravity}= \beta F$) and compute the mass $M$ through other means, e.g., by evaluating the ADM mass (see, e.g., \cite{gibbonshawking}).    The straightforward application requires a static Lorentzian geometry or, equivalently, a Euclidean $U(1)$ invariance reflecting thermodynamic equilibrium.   The metrics (\ref{sph},\ref{Esph}) do not  na\"\i vely satisfy these properties.   In this approach, the entropy arises from the mass and free energy of the spacetime as a whole, and is not evidently associated to the horizon.

\paragraph{(b) Smooth variations -- varying $\beta$ and $M$ simultaneously: }
In this method we want to evaluate the free energy in the canonical ensemble for gravity as a function of the inverse temperature $\beta$ only.  Thus, we vary the mass of the spacetime along with temperature to maintain the appropriate relation between them, and view the resulting regular Euclidean solutions as classical saddle points of the quantum gravitational path integral (see, e.g., \cite{susskind}).   Evaluating the  action on these solutions yields a free energy  $\beta F(\beta)=I_{\rm gravity}$  as a function of the inverse temperature $\beta$ only. Standard thermodynamics then provides the
entropy as:
\begin{equation} \label{entr1}
S= (\beta \partial_{\beta} -1) (\beta F(\beta))
\end{equation}
Written this way, it is evident that any contribution to the Euclidean action that is linear in $\beta$ will not contribute to the entropy.   Since we are evaluating the action on smooth solutions to the vacuum Einstein equations, the only contribution comes from the boundary terms \cite{gibbonshawking}.   If we held the mass fixed while varying $\beta$, these boundary terms at infinity would be proportional to $\beta$ since the geometries would be locally the same for any $\beta$.   But since we are varying the mass $M$ with $\beta$ to keep the geometries regular, the boundary terms and the free energy are not proportional to $\beta$ and give rise to an entropy.  Thought about this way, it is again not obvious that the entropy is associated to the horizon.

To clarify this for planar Rindler and black hole spacetimes, one can split the computation into a contribution from a disc surrounding the origin (i.e. the Euclidean horizon) and an annulus extending from the boundary of the disc to infinity (see \cite{centennial,simonreview}).   Explicit computation then demonstrates that the boundary terms for the annulus (i.e. one at infinity and one at the boundary of the disc) combine to give a contribution that is linear in $\beta$.  Thus  the annulus makes no contribution to the entropy.  What remains is the contribution from the boundary of the disc.  This reproduces the entropy computed above and gives  $S=\mathcal{A}/4G_N$ where $\mathcal{A}$ is the area of the horizon.   By shrinking the radius of the disc to an arbitrarily small size, we see that the entropy is associated to the horizon. The computation can ``transport" the entropy to infinity, since this quantity essentially arises from the topology of the Euclidean saddlepoint.

\paragraph{(c) Conical defects -- Varying $\beta$ keeping the mass fixed: }  Another way to compute the entropy is to  use (\ref{entr1}), but varying only $\beta$ and keeping $M$ fixed. This amounts to changing the proper size of the thermal circle without adjusting anything else in the solution.  This introduces a conical defect/excess at the tip of the Euclidean cigar, which leads to a curvature singularity $4\pi \frac{\delta\beta}{\beta} \delta^2(P)$ localized at the tip of the cigar, i.e. the horizon (see e.g. \cite{fursaevsold,BTZconic,fursaevsnew}).  If we evaluate the Euclidean action on these solutions, the contributions from the bulk of the geometry and the boundary at infinity are necessarily proportional to the periodicity $\beta$, since the geometry is locally  the same for any $\beta$.   Thus, these terms will make no contribution to the entropy after insertion in  the formula (\ref{entr1}).  But the conical singularity makes the additional contribution $ \frac{\delta\beta}{\beta}  \frac{\mathcal{A}}{4G_N}$ to
the Euclidean action in (\ref{entr1}), which once more reproduces the entropy $\mathcal{A}/4G_N$.

What is the justification for varying $\beta$ while keeping the mass fixed, besides
the observation that it reproduces the correct entropy?  One can  understand this roughly as follows. The Euclidean path integral
can be interpreted as computing ${\rm Tr}\,e^{-\beta H}$. If we were doing field theory, not gravity, changing the periodicity of
Euclidean time while keeping spatial slices fixed is equivalent to computing ${\rm Tr}\,e^{-(\beta+\delta\beta)H}$. In gravity it
is not entirely clear that this interpretation is valid, because there is no local Hamiltonian; if there were one, the entropy would
indeed follow from (\ref{entr1}).  Specifically, if we denote $\rho=\exp(-\beta H)$, then (\ref{entr1}) becomes:
\begin{equation}
(\beta\partial_{\beta}-1)\left[- \log {\rm Tr}\, \rho^{1+\frac{\delta\beta}{\beta}}\right] = \left. (1-\partial_{\epsilon})\log {\rm Tr} 
\rho^{1+\epsilon}\right|_{\epsilon=0} = -\left. \partial_{\epsilon} {\rm Tr}\,\hat{\rho}^{1+\epsilon}\right|_{\epsilon=0} =- {\rm Tr}
(\hat{\rho}\log\hat{\rho})
\end{equation}
Here, $\hat{\rho}=\rho/{\rm Tr}\rho$ is the normalized density matrix. Thus, we see that (\ref{entr1}) indeed computes
the entropy when we only vary $\beta$ and this interpretation is correct.

\paragraph{(d) Replica trick}
The replica trick is closely related to the computation we just outlined. Instead of computing ${\rm Tr} 
\rho^{1+\epsilon}$, we compute ${\rm Tr}\rho^n$ for integer $n$ only, and then perform an analytic continuation in $n$.
This is especially useful whenever it is difficult to compute ${\rm Tr} 
\rho^{1+\epsilon}$ for small $\epsilon$ directly. It does, however, assume that ${\rm Tr}\rho^n$ is reasonably well-behaved
and analytic for non-integer $n$. The replica trick can, for example, break down in spin glasses \cite{braymoore} (see the review \cite{spinglassreview}) and in 
systems with spontaneous symmetry breaking, but we are not aware of gravitational examples of either phenomenon.   We will see that for spherical Rindler space the replica approach will be most useful, since the Euclidean continuation (\ref{Esph}) is well defined up to a conical defect for $\tau \sim \tau + 2\pi n$, but not for general periodicities.

\section{Entropy of spherical Rindler space -- near horizon limit}

Recall that the Euclidean spherical Rindler metric (\ref{Esph}) is periodic in imaginary time $\tau$ and that the temporal circle pinches
off smoothly at $r=0$, as it does for black holes.   Based on these observations, 
one might be inclined to associate a temperature to spherical Rindler.
However, the metric is not invariant under translations of $\tau$,  raising the issue of whether
the system is in thermal equilibrium or not, and whether standard thermodynamical relations
are applicable.

When is a system in equilibrium? In  statistical mechanics, we would check this by looking at the 
interface between a system and its heat bath. For a black hole the interface is the horizon: this is evident, because the only way to change the temperature of a black hole is to drop across its horizon an object with a mass comparable to its charges.   This suggests that we can view a gravitational system with a horizon as being in equilibrium with its thermal bath when its near-horizon geometry is the same as that of a static black hole. This criterion, which ensures consistency with the zeroth law of thermodynamics, is satisfied by spherical Rindler space whose near-horizon ($r \to 0$) metric has the leading terms
\begin{equation}
-r^2 dt^2 + dr^2 + R_0^2 d\Omega_{d-1}^2 . \label{near-hor}
\end{equation}
%(see eq.~\ref{near-hor}). 
If the system is in thermal equilibrium, what is then the meaning of the time dependence of the overall geometry? Our view is that it represents some intrinsic dynamics, which does not lead to energy flow
across the horizon and which also does not carry any entropy. This is in line with the standard
intuition, which associates entropy in general relativity only to horizons and views smooth
geometries as coherent states in the underlying microscopic system.  For example, one could imagine a 
gravitational wave, which is deflected but not absorbed by a black hole. Another example,
which is time-dependent but does not carry any entropy, is global de Sitter space.
%  In fact, the $\cosh t$ time dependence of the spherical equal time sections of global de Sitter space is somewhat similar to the time dependence of  spherical Rindler space. 

These considerations lead to a heuristic argument for the entropy of spherical Rindler space. Consider the near-horizon geometry (\ref{near-hor}) of this space and think of it as capturing an equilibrium between a system (``hole'') and a heat bath (exterior), which may or may not itself be in internal equilibrium. Now for the purposes of computing the entropy, any of the methods discussed in Section~\ref{stand} can be applied. All of these methods yield the same answer,
\be
S= \frac{\mathcal{A}}{4 G}\,,
\ee
where $\mathcal{A}$ is the area of a $(d-1)$-sphere of radius $R_0$.

\section{Entropy of spherical Rindler space -- replica method}

If we drop the a priori assumption that entropy originates from the near-horizon region, 
we need a more direct computation of the entropy. Recall that the Euclidean
metric is:
\begin{equation}
ds_E^2 = r^2 d\tau^2 + dr^2 + (R_0 + r \cos \tau)^2 d\Omega_{d-1}^2
\end{equation}
If we ignore the transverse directions, the metric describes the Euclidean plane in polar coordinates. 
The periodicity of the $\tau$-direction is $\beta_0 = 2\pi$, which can be read off both from regularity at $r=0$ and 
from single-valuedness of $g_{\Omega\Omega}$. The center of the polar coordinates is the 
Euclidean continuation of the horizon. Looking  at the transversal directions, 
we see that the size of the sphere goes to zero where $r \cos \tau = - R_0$, so the space caps off 
on this locus. To understand this, change coordinates to $R = R_0 + r \cos \tau$ and $T_E = r \sin \tau$ and 
recognize the result as the analytic continuation of (\ref{radial}). Importantly, even though our 
Lorentzian metric excludes a part of the Minkowski spacetime, the Euclidean continuation does not 
know about this exclusion. There is a topological reason for this: while in Lorentzian 
signature the horizon is a codimension-1 locus that separates two regions in spacetime, 
its Euclidean signature is codimension-2 and can be circumnavigated. In what way, then, 
is the information about the horizon at $R_0$ present in the Euclidean continuation? 
It is there as the choice of center of polar coordinates for the $\tau, r$-plane. This 
choice defines a vector field, which generates time translations. 
%The term $\beta_0 \langle M \rangle$ in eq.~(\ref{legendre}) is defined with respect to this notion of time. 
The Euclidean continuation of the horizon is $r=0$, the locus where $\partial / \partial \tau$ degenerates.

Looking back at the standard methods in Section~\ref{stand}, we see that none of them applies except for
the last one, the replica trick. All the other methods require translational invariance in time and global
thermal equilibrium.   In particular, the direct computation in Method (a)  requires a notion of temperature and mass. It is not clear how the latter quantity should be defined in the absence of time translational invariance,  since the standard ADM definition cannot be applied. 

The smooth variations in Method (b) require one to look at a one-parameter family of smooth solutions as a function of temperature. Although we appear to have a notion of temperature, it cannot be varied while keeping the solution regular. The only free parameter we have is $R_0$, but it seems unrelated to the temperature. It would be interesting to analyze to what extent $R_0$ can be viewed as a thermodynamic variable itself, but for the time being we cannot apply the first method.

The conical defect approach in Method (c) would require a modification of the temperature while keeping all other parameters fixed. In particular, we would  like to keep the geometry of the $t=\tau=0$ slice fixed and then look for  solutions of the field equations with different temperatures. In the standard black hole case this gave rise to the conical deficit solutions. For spherical Rindler this does not work, as the periodicity is  determined both by regularity at $r=0$ as well as the presence of terms containing $\cos\tau$. A Euclidean metric of the form 
\begin{equation}
ds_E^2 = a^2 r^2 d\tau^2 + dr^2 + (R_0 + r \cos b\tau)^2 d\Omega_{d-1}^2 
\end{equation}
is only a solution of the Euclidean field equations if $a^2=b^2$, and is then equivalent to (\ref{Esph})
by a rescaling of the $\tau$ coordinate. There are also no other solutions of the Einstein equations
with the required properties. Spherical symmetry and Birkhoff's theorem imply that the solution would
have to be of the form of the Euclidean Schwarzschild black hole with arbitrary periodicity of
Euclidean time, but none of these are of the form that we are looking for.

The absence of semiclassical saddle points describing the trace of the density matrix $\Tr \rho^{1+\epsilon}$ that appears in Method (c) is quite natural if we are dealing with a system with a time-dependent Hamiltonian $H(\tau)$. Writing $P$ for path-ordering, we have 
\begin{equation}
\rho = P \exp\, \left\{-\!\!\int_0^{\beta} H(\tau) d\tau \right\}
\end{equation}
with $\beta=2\pi$ for (\ref{Esph}), but 
\begin{equation}
\rho^{1+\epsilon} \neq P \exp\, \left\{ -\!\!\int_0^{\beta(1+\epsilon)} H(\tau) d\tau \right\}.
\end{equation}
The entanglement Hamiltonian $\hat{H}$, defined by $\exp(-\beta \hat H)=\rho$, is most
likely a complicated non-local operator, and one does not expect that 
time evolution by $\hat{H}$ is described by
semiclassical saddlepoints for arbitrary time intervals. 

All that remains is  the replica trick in Method (d). This method can still be used, because 
with the $\tau$-periodicity $2\pi n$ ($n \in \mathbb{N}$),
the Euclidean metric (\ref{Esph}) is a proper solution
with a conical defect at the origin. We can once more compute the Euclidean partition function by decomposing the space
in terms of a small disc around the origin plus the remainder. The remainder will be an exact 
$n$-fold copy of the $n=1$ answer, contributing $Z_1^n$ to the partition function. The conical
defect contributes $\exp\big((1-n)\mathcal{A}/4G\big)$ to the partition function. Therefore, 
\begin{equation}
\log {\rm Tr} \rho^n = (1-n) \frac{\mathcal{A}}{4G} + n \log {\rm Tr} \rho.
\end{equation}
This expression has an obvious continuation to non-integer values of $n$. Up to possible subtleties 
associated to the analytic continuation in $n$ mentioned in Section~\ref{stand}, the entropy
is then given by acting with $(1-\partial_n)$ on the above expression and taking $n=1$. The result is once more that 
\be
S=\mathcal{A}/4G \, .
\ee
This derivation complements the heuristic near-horizon argument in the previous section.

\section{Discussion}

We have calculated the gravitational entropy of spherical Rindler space -- a time-dependent spacetime bounded by an acceleration horizon, which is defined by a family of radially accelerating observers. The time dependence is a consequence of the fact that the observers' accelerations are not parallel. We have carried out the calculation in two ways, first using a near-horizon argument and second using the replica trick, each time obtaining $S = \mathcal{A}/4G$ where $\mathcal{A}$ is the area of the horizon.

This may seem surprising. After all, for conventional black holes the large underlying degeneracy and the horizon are  associated to the presence of a mysterious spacetime singularity, whereas here we are simply dealing with a spherical hole in the well-understood, empty flat space.
One answer has been suggested in \cite{mav1, mav2} (see also \cite{swingle1, swingle2}). These papers argue that a necessary condition for a connected, semiclassical spacetime to emerge from the underlying theory of quantum gravity is entanglement. If so, spacetime should be viewed as a geometrization of entanglement in the Hilbert space of quantum gravity microstates, with areas of surfaces computing entanglement entropies connecting complementary subsectors of the theory \cite{myers1, myers2}. Interpreted in this light, the entropy of spherical Rindler space should be viewed as the entanglement entropy between the quantum gravity states describing (the domains of dependence of) the exterior and the interior of a circle of radius $R_0$. If we add a small negative cosmological constant to lift our computation to a holographic setup, the entangled degrees of freedom are, respectively, the ultraviolet and the infrared of the dual field theory. Such entanglement entropy has been computed for a weakly interacting field theory \cite{uvir}, but it is difficult to extend that computation to a strongly coupled regime. Perhaps gravitational computations like the one presented in this paper are the way to do this.

Our computation provides an independent check of the relation between connectedness and entanglement. Recall that ordinary (planar) Rindler wedges come in complementary pairs, which are spacelike separated from one another. Consider an analogous `complementary wedge' to spherical Rindler space -- the region of spacetime, which is spacelike separated from every point in spherical Rindler space. Viewed in spherical coordinates (\ref{radial}), it is the domain of dependence of the $T = 0$ disc $R < R_0$ -- i.e., the `radial causal diamond' $R \pm T < R_0$. Looking back at the definition of the spherical Rindler coordinates (\ref{sphrindler}), we realize that it works equally well for the radial diamond. The only subtlety is that inside the radial diamond the Rindler time runs backwards, much like in the second asymptotic region of the eternal black hole. Because the coordinates covering spherical Rindler space and the radial diamond are related to spherical Minkowski coordinates in the same way, the Euclidean continuation of both complementary `wedges' is the same, so their entropies must be equal!       This is exactly what we would expect of entanglement entropy of complementary regions in a pure state.

The recent paper \cite{maldacena} (see also \cite{fursearlier}) likewise considers the problem of computing the gravitational entropy of a time-dependent spacetime. It follows a similar route to the formal computation presented in this paper, but makes the additional assumption that smooth gravitational saddlepoints can be found for different periods  of Euclidean time. With this assumption the authors of \cite{maldacena} were able to derive the condition that the origin of polar coordinates $r=0$ must be a minimal surface, apparently extending the result of \cite{casinihuertamyers} and fixing the proof \cite{fursaev} of the Ryu-Takayanagi proposal that was criticized in \cite{headrick} (see \cite{rebuttal} for a response). For spherical Rindler space, the horizon is a sphere $R = R_0$ in Minkowski space, so it is not in this sense a minimal surface. Using Ref.~\cite{maldacena}, we can therefore conclude that the free energy at general $\beta$ should not be given by the action evaluated on a regular metric. Indeed, we do not expect this to be the case.  
As we discussed above, the effective entanglement Hamiltonian $\hat{H}$ (defined as the log of the density matrix generated by  translation around Euclidean time) is likely to be a complicated non-local operator, and  time evolution by $\hat{H}$  is unlikely to be described by smooth semiclassical saddlepoints for arbitrary time intervals.

\section*{Acknowledgements} We thank Allan Adams, Fernando Alday, Raphael Bousso, Jos\'e Barb\'on, Jean-Sebastien Caux, Goffredo Chirco, Ben Freivogel, Micha{\l} Heller, Veronika Hubeny, Romuald Janik, Juan Jottar, Klaus Larjo, Matt Lippert, Juan Maldacena, Robert Myers, Tatsuma Nishioka, Carlos N\'u\~nez, Kyriakos Papadodimas, Mark Van Raamsdonk, Mukund Rangamani, Simon Ross, and Erik Verlinde for valuable discussions. This project was initiated at the Holographic Thermalization workshop in Leiden and benefitted from the Black Holes, Horizons and Quantum Information workshop at CERN. VB was supported by DOE grant DE-FG02-05ER-41367 and by the Fondation Pierre-Gilles de Gennes. The work of BDC is supported by the ERC Advanced Grant 268088-EMERGRAV. This work is part of the research programme of the Foundation for Fundamental Research on Matter (FOM), which is part of the Netherlands Organisation for Scientific Research (NWO).


\begin{thebibliography}{..}



\bibitem{rt}
  S.~Ryu and T.~Takayanagi,
  ``Holographic derivation of entanglement entropy from AdS/CFT,''
  Phys.\ Rev.\ Lett.\  {\bf 96}, 181602 (2006)
  [hep-th/0603001].
  %%CITATION = HEP-TH/0603001;%%
  
\bibitem{hrt}
  V.~E.~Hubeny, M.~Rangamani and T.~Takayanagi,
  ``A Covariant holographic entanglement entropy proposal,''
  JHEP {\bf 0707} (2007) 062
  [arXiv:0705.0016 [hep-th]].
  %%CITATION = ARXIV:0705.0016;%%
  
\bibitem{mav1}
  M.~Van Raamsdonk,
  ``Comments on quantum gravity and entanglement,''
  arXiv:0907.2939 [hep-th].
  %%CITATION = ARXIV:0907.2939;%%

\bibitem{mav2}
  M.~Van Raamsdonk,
  ``Building up spacetime with quantum entanglement,''
  Gen.\ Rel.\ Grav.\  {\bf 42}, 2323 (2010)
  [Int.\ J.\ Mod.\ Phys.\ D {\bf 19}, 2429 (2010)]
  [arXiv:1005.3035 [hep-th]].
  %%CITATION = ARXIV:1005.3035;%%

\bibitem{rqg}
  B.~Czech, J.~L.~Karczmarek, F.~Nogueira and M.~Van Raamsdonk,
  ``Rindler Quantum Gravity,''
  Class.\ Quant.\ Grav.\  {\bf 29}, 235025 (2012)
  [arXiv:1206.1323 [hep-th]].
  %%CITATION = ARXIV:1206.1323;%%

\bibitem{pre-amps}
S.~L.~Braunstein, S.~Pirandola and K.~\.Zyczkowski,
  ``Entangled black holes as ciphers of hidden information,''
  Physical Review Letters 110, {\bf 101301} (2013)
  [arXiv:0907.1190 [quant-ph]].
  %%CITATION = ARXIV:0907.1190;%%
  %41 citations counted in INSPIRE as of 25 Sep 2013

\bibitem{firewalls}
  A.~Almheiri, D.~Marolf, J.~Polchinski and J.~Sully,
  ``Black Holes: Complementarity or Firewalls?,''
  JHEP {\bf 1302}, 062 (2013)
  [arXiv:1207.3123 [hep-th]].
  %%CITATION = ARXIV:1207.3123;%%

\bibitem{uvir}
  V.~Balasubramanian, M.~B.~McDermott and M.~Van Raamsdonk,
  ``Momentum-space entanglement and renormalization in quantum field theory,''
  Phys.\ Rev.\ D {\bf 86}, 045014 (2012)
  [arXiv:1108.3568 [hep-th]].
  %%CITATION = ARXIV:1108.3568;%%

\bibitem{maldacena}
A.~Lewkowycz and J.~Maldacena,
  ``Generalized gravitational entropy,''
  JHEP {\bf 1308}, 090 (2013)
  [arXiv:1304.4926 [hep-th]].
  %%CITATION = ARXIV:1304.4926;%%
  %45 citations counted in INSPIRE as of 05 Nov 2013

\bibitem{laflamme}
 R.~Laflamme,
   ``Entropy of a Rindler wedge,''
   Phys.\ Lett.\ B, {\bf 196}, 449 (1987).
 %%CITATION = PHLTA,B196,449;%% 

\bibitem{bekenstein}
  J.~D.~Bekenstein,
  ``Black holes and entropy,''
  Phys.\ Rev.\ D {\bf 7}, 2333 (1973).
  %%CITATION = PHRVA,D7,2333;%%
  
\bibitem{bch}
  J.~M.~Bardeen, B.~Carter and S.~W.~Hawking,
  ``The Four laws of black hole mechanics,''
  Commun.\ Math.\ Phys.\  {\bf 31} (1973) 161.
  %%CITATION = CMPHA,31,161;%%

\bibitem{gibbonshawking}
  G.~W.~Gibbons and S.~W.~Hawking,
  ``Action Integrals and Partition Functions in Quantum Gravity,''
  Phys.\ Rev.\ D {\bf 15}, 2752 (1977).
  %%CITATION = PHRVA,D15,2752;%%
  
\bibitem{centennial}
  S.~W.~Hawking, ``The path-integral approach to quantum gravity,'' in
{\it General Relativity: An Einstein Centenary Survey}, eds. S.W. Hawking
and W. Israel (Cambridge, 1979), Chapter 15.

\bibitem{kallosh}
  R.~Kallosh, T.~Ortin and A.~W.~Peet,
  ``Entropy and action of dilaton black holes,''
  Phys.\ Rev.\ D {\bf 47}, 5400 (1993)
  [hep-th/9211015].
  %%CITATION = HEP-TH/9211015;%%

\bibitem{susskind} 
  L.~Susskind,
  ``Some speculations about black hole entropy in string theory,''
  In *Teitelboim, C. (ed.): The black hole* 118-131
  [hep-th/9309145].
  %%CITATION = HEP-TH/9309145;%%

\bibitem{iyerwald}
  V.~Iyer and R.~M.~Wald,
  ``A Comparison of Noether charge and Euclidean methods for computing the entropy of stationary black holes,''
  Phys.\ Rev.\ D {\bf 52} (1995) 4430
  [gr-qc/9503052].
  %%CITATION = GR-QC/9503052;%%

\bibitem{iyerwald2}
  V.~Iyer and R.~M.~Wald,
  ``Some properties of Noether charge and a proposal for dynamical black hole entropy,''
  Phys.\ Rev.\ D {\bf 50}, 846 (1994)
  [gr-qc/9403028].
  %%CITATION = GR-QC/9403028;%%

  \bibitem{Parattu:2013gwa} 
  K.~Parattu, B.~R.~Majhi and T.~Padmanabhan,
  ``The Structure of the Gravitational Action and its relation with Horizon Thermodynamics and Emergent Gravity Paradigm,''
  arXiv:1303.1535 [gr-qc].
  %%CITATION = ARXIV:1303.1535;%%


\bibitem{simonreview} 
  S.~F.~Ross,
  ``Black hole thermodynamics,''
  hep-th/0502195.
  %%CITATION = HEP-TH/0502195;%%

\bibitem{fursaevsold}
  D.~V.~Fursaev and S.~N.~Solodukhin,
  ``On the description of the Riemannian geometry in the presence of conical defects,''
  Phys.\ Rev.\ D {\bf 52} (1995) 2133
  [hep-th/9501127].
  %%CITATION = HEP-TH/9501127;%%

\bibitem{BTZconic}
  M.~Banados, C.~Teitelboim and J.~Zanelli,
  ``Black hole entropy and the dimensional continuation of the Gauss-Bonnet theorem,''
  Phys.\ Rev.\ Lett.\  {\bf 72} (1994) 957
  [gr-qc/9309026].
  %%CITATION = GR-QC/9309026;%%

\bibitem{fursaevsnew}
  D.~V.~Fursaev, A.~Patrushev and S.~N.~Solodukhin,
  ``Distributional Geometry of Squashed Cones,''
  arXiv:1306.4000 [hep-th].
  %%CITATION = ARXIV:1306.4000;%%
  %5 citations counted in INSPIRE as of 26 Sep 2013

\bibitem{braymoore}
  A.~J.~Bray and M.~A.~Moore,
  ``Replica-Symmetry Breaking in Spin-Glass Theories,''
  Phys.\ Rev.\ Lett.\  {\bf 41}, 1068 (1978).

\bibitem{spinglassreview}
T.~Castellani and A.~Cavagna,
``Spin-glass theory for pedestrians',''
J.\ Stat.\ Mech.\ (2005) P05012, [cond-mat/0505032].

\bibitem{swingle1}
  B.~Swingle,
  ``Entanglement Renormalization and Holography,''
  Phys.\ Rev.\ D {\bf 86}, 065007 (2012)
  [arXiv:0905.1317 [cond-mat.str-el]].
  %%CITATION = ARXIV:0905.1317;%%

\bibitem{swingle2}
  B.~Swingle,
  ``Constructing holographic spacetimes using entanglement renormalization,''
  arXiv:1209.3304 [hep-th].
  %%CITATION = ARXIV:1209.3304;%%

\bibitem{myers1}
  E.~Bianchi and R.~C.~Myers,
  ``On the Architecture of Spacetime Geometry,''
  arXiv:1212.5183 [hep-th].
  %%CITATION = ARXIV:1212.5183;%%

\bibitem{myers2}
  R.~C.~Myers, R.~Pourhasan and M.~Smolkin,
  ``On Spacetime Entanglement,''
  JHEP {\bf 1306}, 013 (2013)
  [arXiv:1304.2030 [hep-th]].
  %%CITATION = ARXIV:1304.2030;%%

\bibitem{fursearlier}
D.~V.~Fursaev,
  ``Entanglement entropy in quantum gravity and the plateau problem,''
  Phys.\ Rev.\ D {\bf 77}, 124002 (2008)
  [arXiv:0711.1221 [hep-th]].
  %%CITATION = ARXIV:0711.1221;%%
  %16 citations counted in INSPIRE as of 25 Sep 2013

\bibitem{casinihuertamyers}
  H.~Casini, M.~Huerta and R.~C.~Myers,
  ``Towards a derivation of holographic entanglement entropy,''
  JHEP {\bf 1105} (2011) 036
  [arXiv:1102.0440 [hep-th]].
  %%CITATION = ARXIV:1102.0440;%%
  %88 citations counted in INSPIRE as of 03 May 2013

\bibitem{fursaev}
D.~V.~Fursaev,
  ``Proof of the holographic formula for entanglement entropy,''
  JHEP {\bf 0609}, 018 (2006)
  [hep-th/0606184].
  %%CITATION = HEP-TH/0606184;%%

\bibitem{headrick}
  M.~Headrick,
  ``Entanglement Renyi entropies in holographic theories,''
  Phys.\ Rev.\ D {\bf 82}, 126010 (2010)
  [arXiv:1006.0047 [hep-th]].
  %%CITATION = ARXIV:1006.0047;%%

\bibitem{rebuttal}
D.~V.~Fursaev,
  ``Entanglement Renyi Entropies in Conformal Field Theories and Holography,''
  JHEP {\bf 1205}, 080 (2012)
  [arXiv:1201.1702 [hep-th]].
  %%CITATION = ARXIV:1201.1702;%%

\end{thebibliography}
\end{document}